\documentclass[aps,prd,onecolumn,groupedaddress,showpacs,nofootinbib,amssymb]{revtex4}
\usepackage[dvips]{graphicx}
\usepackage{amssymb}
\usepackage{amsmath}
\usepackage{graphicx,color}
\usepackage{amsfonts}
\usepackage{bm}
\usepackage{cancel}
\usepackage{comment}
\usepackage{hyperref}
\usepackage{ulem}

\newcommand{\e}{\mathrm{e}}

\allowdisplaybreaks[4]

\begin{document}

\tolerance=5000

\title{Well-defined $f(Q)$ Gravity, Reconstruction of FLRW Spacetime and Unification of Inflation with Dark Energy Epoch}

\author{Shin'ichi~Nojiri$^{1,2}$}\email{nojiri@gravity.phys.nagoya-u.ac.jp}
\author{S.~D.~Odintsov$^{3,4}$}\email{odintsov@ice.csic.es}

\affiliation{ $^{1)}$ Theory Center, IPNS, KEK, 1-1 Oho, Tsukuba, Ibaraki 305-0801, Japan \\
$^{2)}$ Kobayashi-Maskawa Institute for the Origin of Particles and the Universe, Nagoya University, Nagoya 464-8602, Japan \\
$^{3)}$ Institute of Space Sciences (ICE, CSIC) C. Can Magrans s/n, 08193 Barcelona, Spain \\
$^{4)}$ ICREA, Passeig Luis Companys, 23, 08010 Barcelona, Spain
}

\begin{abstract}

We formulate the convenient and mimetic $f(Q)$ gravities in terms of the metric and four scalar fields with 
the $Q$ being the non-metricity scalar. 
As a result, it is shown that the obtained field equations are well-defined. 
By using the field equations, we show how the $f(Q)$ models which realise any given FLRW spacetime may be reconstructed. 
We construct convenient $f(Q)$ gravity and mimetic $f(Q)$ gravity, which describe the inflation and dark energy epochs and even unify the inflation and dark energy. 
Moreover, a radiation-dominated epoch and early dark energy may be easily included in the above theories as is explicitly shown.
It is shown that the inflationary spectral index $n_s$ and the tensor-to-scalar ratio $r$ are consistent with Planck data for some models of $f(Q)$ gravity. 
Corresponding asymptotic forms of $f(Q)$ and mimetic $f(Q)$ are given.
Finally, some remarks on possible reconstruction for spherically symmetric spacetime are given.

\end{abstract}

\maketitle

\section{Introduction\label{SecI}}

General relativity could seem the most successful candidate for classical gravity. 
Nevertheless, there are many unsolved problems like a dark energy problem, H0-tension and the information loss problem caused by a black hole, 
and also the problem of constructing renormalizable quantum gravity. 
In order to solve these problems or to find any clue for the more fundamental theory of gravity, many modified gravity 
theories alternative to general relativity have been proposed and investigated (see reviews \cite{Capozziello:2011et, Faraoni:2010pgm, Cai:2015emx, Nojiri:2010wj, Nojiri:2017ncd}). 

Among the modified gravities, recently, the theories with non-metricity have been actively 
discussed \cite{Nester:1998mp, BeltranJimenez:2018vdo, Runkla:2018xrv, Capozziello:2022tvv}. 
The non-metricity scalar $Q$ is a fundamental geometrical quantity and the connection is a variable independent from the metric in such theories. 
By imposing conditions that the Riemann tensor and torsion tensor vanish, the connection is written by the four scalar 
fields \cite{Blixt:2023kyr, BeltranJimenez:2022azb, Adak:2018vzk, Tomonari:2023wcs}. 
These scalar fields are often chosen so that the connection vanishes. 
As this choice could fix the gauge symmetry of the general covariance, the condition is called the coincident gauge. 
Under such a condition, the only dynamical variable is the metric. 
The theory where the action is linear in $Q$ is equivalent to
Einstein's general relativity because the difference between $Q$ and the scalar curvature $R$ is a total derivative. 
Hence, this theory is called symmetric teleparallel equivalent to general relativity. 

One may consider an extension of the symmetric teleparallel theory, where the action is defined by $f(Q)$, a function of $Q$. 
This theory is an analogue of $f(R)$ gravity or $f(T)$ gravity, where $T$ is a torsion scalar, but it is not equivalent to them. 
Recently there have been presented some studies on the dynamical degrees of freedom (DOF) 
of $f(Q)$ gravity \cite{Hu:2022anq, DAmbrosio:2023asf, Heisenberg:2023lru, Paliathanasis:2023pqp, Dimakis:2021gby, Hu:2023gui}.
However, the problem has not been completely solved although the only propagating mode in the flat background is a graviton as in Einstein's theory \cite{Capozziello:2024vix}. 
This situation is similar to $f(T)$ gravity \cite{Bamba:2013ooa} but different from that in the $f(R)$ gravity, where extra scalar mode, which appears as a scale of the metric 
of the scalar curvature, propagates. 
In \cite{Hu:2023gui}, it has been shown that the corresponding scalar mode in the $f(Q)$ gravity does not propagate due to the constraint, which is consistent with the 
result in \cite{Capozziello:2024vix}. 

In this paper, we define $f(Q)$ gravity theory by using the metric and only four scalar fields which give the connection as an independent field. 
Although the number of dynamical degrees of freedom on-shell is still not clear, the equations given by the variation of the action 
with respect to these fields become well-defined. 
Note that the functional degrees of freedom in the connection are restricted by the curvature-free and torsion-free conditions 
but it is not so clear what could be valid in the equations given by the variation with respect to the connection. 
Our formulation avoids this problem. 

By using the above formulation, we consider the ``reconstruction'' of the model realising an arbitrary given Friedmann-Lema\^{i}tre-Robertson-Walker (FLRW) spacetime. 
Usually solving the equations for a given gravity model, we find the structure of spacetime as the solution. 
We are interested here in the inverse problem, that is, to find a model that realises the geometry desired from the theoretical and/or observational viewpoints. 
If the model can consistently realise the spacetime, the model can be realistic. 
We often call such systematic formulation the ``reconstruction''. 
In the case of the FLRW spacetime, the consistent formulation of the reconstruction has been well-studied for the scalar-tensor theories \cite{Nojiri:2005pu, Capozziello:2005tf}, 
the Einstein--scalar--Gauss-Bonnet gravity \cite{Nojiri:2006je, Nojiri:2005jg}, and $F (\mathcal{G})$ gravity \cite{Cognola:2006eg, Nojiri:2019dwl}, 
where $\mathcal{G}$ is the Gauss-Bonnet topological invariant, and $f(R)$ gravity \cite{Nojiri:2006gh, Nojiri:2009kx}. 
Some attempts for reconstruction in $f(Q)$ gravity have been considered in \cite{Capozziello:2022wgl, Gadbail:2023klq, Gadbail:2023mvu} and also in \cite{Kaczmarek:2024yju} by using the mimetic formulation. 
In this paper, we give a systematic and versatile formulation. 
By using this formulation, we construct $f(Q)$ gravity realising the inflation epoch and the dark energy epoch, and even unifying both the inflation and the dark energy epochs. 

The paper is composed as follows. 
In the next section, the review of the consistent formulation of $f(Q)$ gravity is given in detail. 
Section~\ref{SecIII} is devoted to the formulation of the FLRW spacetime in $f(Q)$ gravity. 
The consistent and convenient reconstruction of $f(Q)$ as well as of mimetic $f(Q)$ from the arbitrary FLRW universe is given in Section~\ref{SecIV}. 
Inflation, as well as dark energy for such theories, is studied in Section~\ref{SecV}. 
A model of $f(Q)$ gravity which successfully describes inflation with subsequent transition then to a radiation-dominated epoch is constructed. 
Convenient and mimetic $f(Q)$ gravities which describe the dark energy epoch are explicitly constructed in this section, too.
Finally, the unification of inflation with dark energy is realised for convenient and mimetic $f(Q)$ gravities.
The last section is devoted to summary and discussion. 

\section{Brief review of $f(Q)$ gravity\label{SecII}}

We now consider $f(Q)$ gravity, whose formulation is given as the following. 
In previous papers, the treatments of the $f(Q)$ gravity are not so well-defined, which is related to the degrees of freedom. 
For example, it is often used the variation of the action with respect to the connection to obtain the corresponding equations. 
In the case of $f(Q)$ gravity, however, the connection is very restricted by requiring that the torsion and the Riemann tensor should vanish. 
Therefore not all the components of the connection are independent. 
One way to solve this problem is to impose the constraints on the connection, that is, the vanishing torsion and the vanishing Riemann curvature, 
by the Lagrange multiplier fields. Unfortunately, we have not found any paper where the multiplier fields are solved but often the reduced equations are used. 
In this paper, the $f(Q)$ model is defined only by using the metric and four scalar fields $\xi^a$, whose structure will be explained later. 
Therefore, the model under consideration is well-defined. 
The inconsistency in choosing the so-called coincident gauge in the FLRW spacetime can be obtained from the equations given by the variation of the action 
with respect to $\xi^a$ as we will see in the next section. 

The general affine connection on a manifold that is both parallelisable and differentiable can be expressed as follows:
\begin{align}
\label{affine}
{\Gamma^\sigma}_{\mu \nu}= {{\tilde \Gamma}^\sigma}_{\mu \nu} + K^\sigma_{\;\mu \nu} + L^\sigma_{\;\mu \nu}\,.
\end{align}
Here $\tilde \Gamma^\sigma_{\;\mu \nu}$ is the Levi-Civita connection given by the metric,
\begin{align}
\label{Levi-Civita}
{{\tilde\Gamma}^\sigma}_{\mu \nu} = \frac{1}{2} g^{\sigma \rho} \left( \partial_\mu g_{\rho \nu} + \partial_\nu g_{\rho \mu}- \partial_\rho g_{\mu \nu}\right)\,.
\end{align}
Furthermore, ${K^\sigma}_{\mu \nu}$ represents the contortion which is defined by using the torsion tensor
${T^\sigma}_{\mu \nu}={\Gamma^\sigma}_{\mu \nu} - {\Gamma^\sigma}_{\nu \mu}$ as follows, 
\begin{align}
\label{contortion}
{K^\sigma}_{\mu \nu}= \frac{1}{2} \left( {T^\sigma}_{\mu \nu} + T^{\ \sigma}_{\mu\ \nu} + T^{\ \sigma}_{\nu\ \mu} \right) \, .
\end{align}
Finally, ${L^\sigma}_{\mu \nu}$ denotes the deformation and is expressed as follows:
\begin{align}
\label{deformation}
{L^\sigma}_{\mu \nu}= \frac{1}{2} \left( Q^\sigma_{\;\mu \nu} - Q^{\ \sigma}_{\mu\ \nu} - Q^{\ \sigma}_{\nu\ \mu} \right)\,.
\end{align}
Here ${Q^\sigma}_{\mu \nu}$ represents the non-metricity tensor expressed as,
\begin{align}
\label{non-metricity}
Q_{\sigma \mu \nu}= \nabla_\sigma g_{\mu \nu}= \partial_\sigma g_{\mu \nu} - {\Gamma^\rho}_{\sigma \mu } g_{\nu \rho} - {\Gamma^\rho}_{\sigma \nu } g_{\mu \rho } \,.
\end{align}
We use $Q_{\sigma \mu \nu}$ in order to construct the $f(Q)$ gravity. 

In the following, we consider the case without torsion, therefore we assume ${\Gamma^\sigma}_{\mu\nu} = {\Gamma^\sigma}_{\nu\mu}$. 
Symmetric teleparallel theories of gravity are obtained by requiring the Riemann tensor to vanish, 
\begin{align}
\label{curvatures}
R^\lambda_{\ \mu\rho\nu} \equiv \Gamma^\lambda_{\mu\nu,\rho} -\Gamma^\lambda_{\mu\rho,\nu} + \Gamma^\eta_{\mu\nu}\Gamma^\lambda_{\rho\eta}
 - \Gamma^\eta_{\mu\rho}\Gamma^\lambda_{\nu\eta} =0 \, ,
\end{align}
whose solution is given by using four fields $\xi^a$ $\left( a = 0,1,2,3 \right)$ as follows, 
\begin{align}
\label{G1B}
{\Gamma^\rho}_{\mu\nu}=\frac{\partial x^\rho}{\partial \xi^a} \partial_\mu \partial_\nu \xi^a \, .
\end{align}
As we show soon, $\xi^a$'s should be scalar fields and we may regard $e^a_\mu\equiv \partial_\mu \xi^a$'s as vierbein fields. 
There is a gauge symmetry of the general coordinate transformation, so we often choose the gauge condition ${\Gamma^\rho}_{\mu\nu}=0$, which is called 
the coincident gauge and can be realised by choosing $\xi^a=x^a$. 
The gauge condition, however, often contradicts the FLRW universe and the spherically symmetric spacetime. 

When we consider the infinitesimal transformation,
\begin{align}
\label{G2}
\xi^a \to \xi^a + \delta \xi^a \, .
\end{align}
one finds
\begin{align}
\label{G1}
\Gamma^\rho_{\mu\nu} \to \Gamma^\rho_{\mu\nu} + \delta \Gamma^\rho_{\mu\nu}
\equiv \Gamma^\rho_{\mu\nu} - \frac{\partial x^\rho}{\partial \xi^a} \partial_\sigma \delta\xi^a\frac{\partial x^\sigma}{\partial \xi^b}\partial_\mu \partial_\nu \xi^b
+ \frac{\partial x^\rho}{\partial \xi^a} \partial_\mu \partial_\nu \delta \xi^a \, , 
\end{align}
which can be used later to find the field equations. 
By regarding $\xi^a$'s as scalar fields, under the coordinate transformation $x^\mu\to x^\mu + \epsilon^\mu$ with infinitesimally small functions $\epsilon^\mu$, 
we find $\delta\xi^a = \epsilon^\mu \partial_\mu \xi^a$ and therefore 
\begin{align}
\label{G1GCT}
\delta \Gamma^\rho_{\mu\nu}
=&\, \epsilon^\sigma \partial_\sigma \Gamma^\rho_{\mu\nu} 
 - \partial_\sigma \epsilon^\eta \frac{\partial x^\rho}{\partial \xi^a} \partial_\eta \xi^a \frac{\partial x^\sigma}{\partial \xi^b} \partial_\mu \partial_\nu \xi^b 
+ \frac{\partial x^\rho}{\partial \xi^a} 
\left( \partial_\mu \epsilon^\eta \partial_\nu \partial_\eta \xi^a + \partial_\nu \epsilon^\eta \partial_\mu \partial_\eta \xi^a + \partial_\mu \partial_\nu \epsilon^\eta \partial_\eta \xi^a 
\right) \nonumber \\
=&\, \epsilon^\sigma \partial_\sigma \Gamma^\rho_{\mu\nu} - \partial_\sigma \epsilon^\rho \Gamma^\sigma_{\mu\nu} 
+ \partial_\mu \epsilon^\eta \Gamma^\rho_{\eta\nu} + \partial_\nu \epsilon^\eta \Gamma^\rho_{\mu\eta} 
+ \partial_\mu \partial_\nu \epsilon^\rho \, , 
\end{align}
where the last term is nothing but the inhomogeneous term, which guarantees the general covariance of the covariant derivative. 
This also shows that $\xi^a$'s are not vector fields but scalar fields. 

In the symmetric teleparallel theory, we use the non-metricity tensor in (\ref{non-metricity}) and the scalar of the non-metricity defined as follows, 
\begin{align}
\label{non-m scalar}
Q\equiv g^{\mu \nu} \left( {L^\alpha}_{\beta \nu}{L^\beta}_{\mu \alpha} - {L^\beta}_{\alpha \beta} {L^\alpha}_{\mu \nu} \right)
 -Q_{\sigma \mu \nu} P^{\sigma \mu \nu} \,.
\end{align}
Here ${L^\sigma}_{\mu \nu}$ and $P^{\sigma \mu \nu}$ are defined by,
\begin{align}
\label{deformationB}
{L^\sigma}_{\mu \nu}\equiv &\, \frac{1}{2} \left( Q^\sigma_{\;\mu \nu} - Q^{\ \sigma}_{\mu\ \nu} - Q^{\ \sigma}_{\nu\ \mu} \right)\, , \\
\label{non-m conjugate}
{P^\sigma}_{\mu \nu} \equiv &\, \frac{1}{4} \left\{ - {Q^\sigma}_{\mu \nu} + Q^{\ \sigma}_{\mu\ \nu} + Q^{\ \sigma}_{\nu\ \mu}
+ Q^\sigma g_{\mu \nu}- \tilde{Q}^\sigma g_{\mu \nu} - \frac{1}{2} \left( {\delta^\sigma}_\mu Q_\nu + {\delta^\sigma}_\nu Q_\mu \right) \right\}\, ,
\end{align}
and $Q_\sigma$ and $\tilde{Q}_\sigma$ are defined as
$Q_\sigma \equiv Q^{\ \mu}_{\sigma\ \mu}$ and $\tilde{Q}_\sigma=Q^\mu_{\ \sigma \mu}$. 
Then one gets
\begin{align}
\label{Q}
Q=&\, - \frac{1}{4} g^{\alpha\mu} g^{\beta\nu} g^{\gamma\rho} \nabla_\alpha g_{\beta\gamma} \nabla_\mu g_{\nu\rho}
+ \frac{1}{2} g^{\alpha\mu} g^{\beta\nu} g^{\gamma\rho} \nabla_\alpha g_{\beta\gamma} \nabla_\rho g_{\nu\mu}
+ \frac{1}{4} g^{\alpha\mu} g^{\beta\gamma} g^{\nu\rho} \nabla_\alpha g_{\beta\gamma} \nabla_\mu g_{\nu\rho} \nonumber \\
&\, - \frac{1}{2} g^{\alpha\mu} g^{\beta\gamma} g^{\nu\rho} \nabla_\alpha g_{\beta\gamma} \nabla_\nu g_{\mu\rho} \, .
\end{align}
The action of $f(Q)$ gravity is given by 
\begin{align}
\label{ActionQ}
S=\int d^4 x \sqrt{-g} f(Q)\, .
\end{align}
In the following, we regard the metric $g_{\mu\nu}$ and $\xi^a$ as independent fields. 
Then it follows
\begin{align}
\label{eq1}
\mathcal{G}_{\mu\nu} \equiv &\, \frac{1}{\sqrt{-g}} g_{\mu\rho} g_{\nu\sigma}\frac{\delta S}{\delta g_{\rho\sigma}} \nonumber \\
=&\, \frac{1}{2} g_{\mu\nu} f(Q)
 - f'(Q) g^{\alpha\beta} g^{\gamma\rho} \left\{ -\frac{1}{4} \nabla_\mu g_{\alpha\gamma} \nabla_\nu g_{\beta\rho}
 - \frac{1}{2} \nabla_\alpha g_{\mu\gamma} \nabla_\beta g_{\nu\rho} \right. \nonumber \\
&\, + \frac{1}{2} \left( \nabla_\mu g_{\alpha\gamma} \nabla_\rho g_{\beta\nu}
+ \nabla_\nu g_{\alpha\gamma} \nabla_\rho g_{\beta\mu} \right)
+ \frac{1}{2} \nabla_\alpha g_{\mu\gamma} \nabla_\rho g_{\nu\beta}
+ \frac{1}{4} \nabla_\mu g_{\alpha\beta} \nabla_\nu g_{\gamma\rho}
+ \frac{1}{2} \nabla_\alpha g_{\mu\nu} \nabla_\beta g_{\gamma\rho} \nonumber \\
&\, \left. - \frac{1}{4} \left( \nabla_\mu g_{\alpha\beta} \nabla_\gamma g_{\nu\rho} + \nabla_\nu g_{\alpha\beta} \nabla_\gamma g_{\mu\rho} \right)
 - \frac{1}{2} \nabla_\alpha g_{\mu\nu} \nabla_\gamma g_{\beta\rho}
 - \frac{1} {4} \left(\nabla_\alpha g_{\gamma\rho} \nabla_\mu g_{\beta\nu} + \nabla_\alpha g_{\gamma\rho} \nabla_\nu g_{\beta\mu} \right)
\right\} \nonumber \\
&\, - \frac{g_{\mu\rho} g_{\nu\sigma}}{\sqrt{-g}} \partial_\alpha \left[ \sqrt{-g} f'(Q) \left\{ - \frac{1}{2} g^{\alpha\beta} g^{\gamma\rho} g^{\sigma\tau} \nabla_\beta g_{\gamma\tau}
+ \frac{1}{2} g^{\alpha\beta} g^{\gamma\rho} g^{\sigma\tau} \left( \nabla_\tau g_{\gamma\beta} + \nabla_\gamma g_{\tau\beta} \right) \right. \right. \nonumber \\
&\, \left. \left. + \frac{1}{2} g^{\alpha\beta} g^{\rho\sigma} g^{\gamma\tau} \nabla_\beta g_{\gamma\tau}
 - \frac{1}{2} g^{\alpha\beta} g^{\rho\sigma} g^{\gamma\tau} \nabla_\gamma g_{\beta\tau}
 - \frac{1}{4} \left( g^{\alpha\sigma} g^{\gamma\tau} g^{\beta\rho} + g^{\alpha\rho} g^{\gamma\tau} g^{\beta\sigma}
\right) \nabla_\beta g_{\gamma\tau}
\right\} \right] \nonumber \\
&\, - \frac{g_{\mu\rho} g_{\nu\sigma}}{\sqrt{-g}} {\Gamma^\rho}_{\alpha\eta} \left[ \sqrt{-g} f'(Q) \left\{ - \frac{1}{2} g^{\alpha\beta} g^{\gamma\eta} g^{\sigma\tau} \nabla_\beta g_{\gamma\tau}
+ \frac{1}{2} g^{\alpha\beta} g^{\gamma\eta} g^{\sigma\tau} \left( \nabla_\tau g_{\gamma\beta} + \nabla_\gamma g_{\tau\beta} \right) \right. \right. \nonumber \\
&\, \left. \left. + \frac{1}{2} g^{\alpha\beta} g^{\eta\sigma} g^{\gamma\tau} \nabla_\beta g_{\gamma\tau}
 - \frac{1}{2} g^{\alpha\beta} g^{\eta\sigma} g^{\gamma\tau} \nabla_\gamma g_{\beta\tau}
 - \frac{1}{4} \left( g^{\alpha\sigma} g^{\gamma\tau} g^{\beta\eta} + g^{\alpha\eta} g^{\gamma\tau} g^{\beta\sigma}
\right) \nabla_\beta g_{\gamma\tau}
\right\} \right] \nonumber \\
&\, - \frac{g_{\mu\rho} g_{\nu\sigma}}{\sqrt{-g}} {\Gamma^\sigma}_{\alpha\eta} \left[ \sqrt{-g} f'(Q) \left\{ - \frac{1}{2} g^{\alpha\beta} g^{\gamma\rho} g^{\eta\tau} \nabla_\beta g_{\gamma\tau}
+ \frac{1}{2} g^{\alpha\beta} g^{\gamma\rho} g^{\eta\tau} \left( \nabla_\tau g_{\gamma\beta} + \nabla_\gamma g_{\tau\beta} \right) \right. \right. \nonumber \\
&\, \left. \left. + \frac{1}{2} g^{\alpha\beta} g^{\rho\eta} g^{\gamma\tau} \nabla_\beta g_{\gamma\tau}
 - \frac{1}{2} g^{\alpha\beta} g^{\rho\eta} g^{\gamma\tau} \nabla_\gamma g_{\beta\tau}
 - \frac{1}{4} \left( g^{\alpha\eta} g^{\gamma\tau} g^{\beta\rho} + g^{\alpha\rho} g^{\gamma\tau} g^{\beta\eta}
\right) \nabla_\beta g_{\gamma\tau}
\right\} \right] \, , \\
\label{eq2}
\mathcal{H}^{\xi\zeta}_\eta \equiv&\, \frac{1}{\sqrt{-g}} \frac{\delta S}{\delta {\Gamma^\eta}_{\xi\zeta}} \nonumber \\
=&\, - f'(Q) \left( 
 - \frac{1}{2} g^{\xi\rho} g^{\zeta\nu} g^{\gamma\mu} 
 - \frac{1}{2} g^{\xi\rho} g^{\gamma\nu} g^{\zeta\mu} 
+ 2 g^{\xi\mu} g^{\zeta\nu} g^{\gamma\rho} \right. \nonumber \\
&\, \qquad \qquad \quad \left. + \frac{1}{4} g^{\xi\mu} g^{\zeta\gamma} g^{\nu\rho} 
 - g^{\xi\nu} g^{\zeta\gamma} g^{\mu\rho} 
 - \frac{1}{2} g^{\mu\zeta} g^{\nu\rho} g^{\xi\gamma} 
 - \frac{1}{2} g^{\mu\gamma} g^{\nu\rho} g^{\xi\zeta} 
\right) g_{\eta\gamma} \nabla_\mu g_{\nu\rho} \, .
\end{align}
By using (\ref{G1}), we find 
\begin{align}
\label{eq3}
X_a \equiv&\, \frac{1}{\sqrt{-g}} \frac{\delta S}{\delta \xi^a} 
= \partial_\sigma
\left\{ \frac{\partial x^\eta}{\partial \xi^a} \frac{\partial x^\sigma}{\partial \xi^b}\partial_\xi \partial_\zeta \xi^b 
\left( \sqrt{-g} \mathcal{H}^{\xi\zeta}_\eta \right) \right\}
+ \partial_\xi \partial_\zeta \left\{ \frac{\partial x^\eta}{\partial \xi^a} \left( \sqrt{-g} \mathcal{H}^{\xi\zeta}_\eta \right) \right\} 
\nonumber \\
=&\, - \partial_\sigma
\left[ \frac{\partial x^\eta}{\partial \xi^a} \frac{\partial x^\sigma}{\partial \xi^b}\partial_\xi \partial_\zeta \xi^b \left\{ \sqrt{-g} f'(Q) 
 \left( 
 - \frac{1}{2} g^{\xi\rho} g^{\zeta\nu} g^{\gamma\mu} 
 - \frac{1}{2} g^{\xi\rho} g^{\gamma\nu} g^{\zeta\mu} 
+ 4 g^{\xi\mu} g^{\zeta\nu} g^{\gamma\rho} \right. \right. \right. \nonumber \\
&\, \qquad \qquad \quad \left. \left. \left. + \frac{1}{4} g^{\xi\mu} g^{\zeta\gamma} g^{\nu\rho} 
 - g^{\xi\nu} g^{\zeta\gamma} g^{\mu\rho} 
 - g^{\mu\zeta} g^{\nu\rho} g^{\xi\gamma} 
 - g^{\mu\gamma} g^{\nu\rho} g^{\xi\zeta} 
\right) 
g_{\eta\gamma} \nabla_\mu g_{\nu\rho} \right\} \right] \nonumber \\
&\, - \partial_\xi \partial_\zeta \left[\frac{\partial x^\eta}{\partial \xi^a} \left\{ \sqrt{-g} f'(Q) 
 \left( 
 - \frac{1}{2} g^{\xi\rho} g^{\zeta\nu} g^{\gamma\mu} 
 - \frac{1}{2} g^{\xi\rho} g^{\gamma\nu} g^{\zeta\mu} 
+ 4 g^{\xi\mu} g^{\zeta\nu} g^{\gamma\rho} \right. \right. \right. \nonumber \\
&\, \qquad \qquad \quad \left. \left. \left. + \frac{1}{4} g^{\xi\mu} g^{\zeta\gamma} g^{\nu\rho} 
 - g^{\xi\nu} g^{\zeta\gamma} g^{\mu\rho} 
 - g^{\mu\zeta} g^{\nu\rho} g^{\xi\gamma} 
 - g^{\mu\gamma} g^{\nu\rho} g^{\xi\zeta} 
\right) 
g_{\eta\gamma} \nabla_\mu g_{\nu\rho} \right\} \right]
\, .
\end{align}
The above quantities are used when we consider the field equations. 

\section{FLRW spacetime in $f(Q)$ gravity\label{SecIII}}

We now consider the spatially flat FLRW spacetime whose metric is given by the following line element, 
\begin{align}
\label{FLRW}
ds^2 = - dt^2 + a(t)^2 \sum_{i=1,2,3} \left( dx^i \right)^2 \, .
\end{align}
We also assume
\begin{align}
\label{xi}
\xi^0 = b(t)\, , \quad \xi^i = x^i\, , 
\end{align}
which gives
\begin{align}
\label{connctn}
{\Gamma^0}_{00}=\gamma(t) \equiv \frac{\ddot b(t)}{\dot b(t)} \, , \quad \mbox{others}=0\, ,
\end{align}
and therefore we obtain, 
\begin{align}
\label{cov}
\nabla_0 g_{00} = 2 \gamma(t)\, , \quad 
\nabla_0 g_{ij} = 2 a \dot a \delta_{ij}\, , \quad \mbox{others}=0 \, .
\end{align}
Eq.~(\ref{xi}) could be a unique choice when we require the invariances of the spacetime under the translation and the rotation. 
Then the metricity scalar in (\ref{Q}) has the following form, 
\begin{align}
\label{QFLRW}
Q = - 6 H^2 \, .
\end{align}
Here the Hubble rate $H$ is defined by $H=\frac{\dot a}{a}$. 
We should note that $Q$ does not contain $\gamma$. 
It is also noted that the difference between $\sqrt{-g}Q$ and the scalar curvature $\sqrt{-g} \tilde R = a^3 \left(6\dot H + 12 H^2\right)$ is given by 
\begin{align}
\label{def}
\sqrt{-g} \left( \tilde R - Q \right) = a^3 \left( 6\dot H + 18 H^2 \right) = \frac{d}{dt} \left( 6 a^3 H \right) \, ,
\end{align}
which is consistently a total derivative. 

By using (\ref{eq1}), one gets
\begin{align}
\label{G00}
\mathcal{G}_{00} =&\, - \frac{1}{2} f(Q) - 6 H^2 f'(Q) \, , \\
\label{Gij}
\mathcal{G}_{ij} 
=&\, a^2 \delta_{ij} \left\{ \frac{f(Q)}{2} + 2 a^{-3} \frac{d}{dt} \left( a^3 H f'(Q) \right)\right\}
\, , \\
\label{Gi0}
\mathcal{G}_{0i}=&\, \mathcal{G}_{i0} 
= 0\, , \\
\label{H000}
\mathcal{H}_0^{00} =&\, \frac{3}{2} f'(Q) \left( \gamma - H \right) \, , \\
\label{Hij0}
\mathcal{H}_0^{ij} =&\, - f'(Q) a^{-2} \delta^{ij} \left( - 2 H + \gamma \right) \, , \\
\label{0ij}
\mathcal{H}_j^{0i} =&\, \frac{f'(Q)}{2} \delta_i^{\ j} \left( 7 H + 5 \gamma \right) \, , 
\end{align}
and (\ref{eq3}) tells
\begin{align}
\label{Xa}
X_0 = \frac{d}{dt} \left[ \frac{\gamma}{b'} \left\{ a^3 \frac{3}{2} f'(Q) \left( \gamma - H \right) \right\} \right] 
+ \frac{d^2}{dt^2} \left\{ \frac{a^3}{b'} f'(Q) \left( \gamma - H \right) \right\} \, , \quad
X_i = 0 \, .
\end{align}
If the connection does not couple directly with the matter, we need to require $X_a=0$, which gives 
\begin{align}
\label{gammaH}
\gamma=H \, ,
\end{align}
what tells that the coincident gauge, where $\gamma=0$, is not consistent with the FLRW spacetime. 

Because $Q$ can be written as $Q=\tilde R+\mbox{total derivative terms}$ ($\tilde R$ is the scalar curvature in Einstein's gravity), 
the model reduces to Einstein's gravity when $f(Q)= \frac{Q}{2\kappa^2}$. 
In fact, Eqs.~(\ref{G00}) and (\ref{Gij}) give 
\begin{align}
\label{Einstein}
\mathcal{G}_{00}= \frac{3}{2\kappa^2}H^2 - \frac{3}{\kappa^2}H^2 = - \frac{3}{2\kappa^2}H^2 \, , \quad 
\mathcal{G}_{ij}= a^2 \delta_{ij} \left( - \frac{3}{2\kappa^2}H^2 +\frac{3}{\kappa^2}H^2 + \frac{1}{\kappa^2} \dot H \right) 
= \frac{1}{2\kappa^2} \left( 3 H^2 + 2 \dot H \right) \, .
\end{align}
Therefore, the standard FLRW equations are reproduced as expected. 

\section{Cosmological Reconstruction of $f(Q)$ Gravity\label{SecIV}}

We now consider how one can reconstruct the realistic universe expansion in the FLRW spacetime (\ref{FLRW}). 
One should note that the reconstruction itself is not so difficult. 
Still, in this paper, we give $f(Q)$ in the form of the integration for arbitrary universe expansion. 
Then one can apply the obtained expression to any expansion. 
We also consider the mimetic theory of $f(Q)$ gravity and the reconstruction in this theory. 
We will give a simple equation, that can be applied to the arbitrary FLRW universe expansion straightforwardly. 

\subsection{Reconstruction by using the form of $f(Q)$}

The equations corresponding to the FLRW equation in General relativity are given by 
\begin{align}
\label{vQ_FLRW0}
- \rho =&\, \mathcal{G}_{00} = - \frac{f(Q)}{2} - 6 H^2 f'(Q) \, , \nonumber \\
 - p a^2 \delta_{ij} =&\, \mathcal{G}_{ij} = a^2 \delta_{ij} \left\{ \frac{f(Q)}{2} + 2 a^{-3} \frac{d}{dt} \left( a^3 H f'(Q) \right)\right\} \nonumber \\
=&\, a^2 \delta_{ij} \left\{ \frac{f(Q)}{2} + 6H^2 f'(Q) + 2 \dot H f'(Q) - 24 H^2 \dot H f''(Q) \right\} \, .
\end{align}
Here $\rho$ and $p$ are the energy density and the pressure of matter, respectively. 
We have also used (\ref{QFLRW}), (\ref{G00}), and (\ref{Gij}). 

One may check the conservation law of the matter, 
\begin{align}
\label{cons}
\dot\rho + 3 H \left( \rho + p \right) = 0 \, .
\end{align}

Note that the conservation of matter can be also obtained only from the matter equation of motion. 
The conservation law of the matter energy-momentum tensor $T_{\mu\nu}$ holds regardless of what theory is considered: Einstein's gravity, $f(Q)$ theory, $f(T)$ or other theory. 
Therefore the conservation should be described by the Levi-Civita connection $\tilde \Gamma^\sigma_{\;\mu \nu}$ in (\ref{Levi-Civita}) for Einstein's gravity 
even if we consider the $f(Q)$ gravity theory, 
\begin{align}
\label{cons2}
0 = \tilde\nabla^\mu T_{\mu\nu} = g^{\rho\mu} \left( \partial_\rho T_{\mu\nu} - \tilde \Gamma^\sigma_{\; \rho\mu} T_{\sigma\nu} 
 - \tilde \Gamma^\sigma_{\; \rho\nu} T_{\sigma\mu} \right) \, .
\end{align}
The equation corresponding to the Einstein equation is given by 
\begin{align}
\label{EQeq}
0= \mathcal{G}_{\mu\nu} + T_{\mu\nu}\, ,
\end{align}
and we are considering the theory with action, the consistency of the Lagrangian theory, which is a condition of functional integrability shows us, 
\begin{align}
\label{QBianchi}
0 = \tilde\nabla^\mu \mathcal{G}_{\mu\nu} \, ,
\end{align}
This is the generalized Bianchi identity in $f(Q)$ gravity theory. 

In the case of $f(Q)$ gravity, there is an exceptional model, where arbitrary development of the universe expansion in the FLRW spacetime (\ref{FLRW}) is a solution. 
When we neglect the contribution of the matter, 
\begin{align}
\label{vQ_FLRW0vacuum}
0 =&\, \mathcal{G}_{00} = - \frac{f(Q)}{2} - 6 H^2 f'(Q) \, , \nonumber \\
0 =&\, \mathcal{G}_{ij} = 
a^2 \delta_{ij} \left\{ \frac{f(Q)}{2} + 6H^2 f'(Q) + 2 \dot H f'(Q) + 24 H^2 \dot H f''(Q) \right\} \, .
\end{align}
Eq.~(\ref{cons}) also tells that in the vacuum, where $\rho=p=0$, the second equation in (\ref{vQ_FLRW0vacuum}) can be obtained from the first equation. 
Therefore one may forget the second equation. 
We should note that by using Eq.~(\ref{QFLRW}), the first equation in (\ref{vQ_FLRW0vacuum}) can be rewritten as 
\begin{align}
\label{vQ_FLRW1}
0 = - \frac{1}{2} f(Q) + Q f'(Q) \, .
\end{align}
Then we find if Q is given by 
\begin{align}
\label{vQ_FLRW2}
f(Q) = f_0 \sqrt{ - Q } \, .
\end{align}
Hence, if we consider the model given by (\ref{vQ_FLRW2}), arbitrary development of the universe expansion is a solution. 
This also shows that when there is no matter, we find $Q=f(Q)=0$ or $f(Q)$ in (\ref{vQ_FLRW2}). 
Therefore at the late time, when the density of matter decreases, and if there is no cosmological constant, 
the spacetime becomes static and flat or $f(Q)$ is asymptotically given by (\ref{vQ_FLRW2}). 

Let us consider the case that there are several kinds of matter. 
We assume the matters satisfy the conservation law $\dot\rho + 3 H \left( \rho + p \right)$ as in (\ref{cons}). 
The conservation law determines the scale dependence of $\rho$ and $p$, $\rho=\rho(a)$ and $p=p(a)$. 
If the scale factor $a=a(t)$ is given, we find the $t$ dependences of $\rho$, $p$, and $Q$, $\rho=\rho(t)$, $p=p(t)$, and $Q=-6H^2 = Q(t)$. 
By solving $Q=Q(t)$ with respect to $t$, $t=t(Q)$, and by substituting the obtained expression into $\rho=\rho(t)$ and $p=p(t)$, 
we find $Q$ dependence of $\rho$ and $p$ as $\rho=\rho(Q)$ and $p=p(Q)$. 
By writing the first equation in (\ref{vQ_FLRW0}) as follows
\begin{align}
\label{vQ_FLRW0Q}
- \rho (Q) = - \frac{f(Q)}{2} + Q f'(Q) = - \sqrt{-Q} \frac{d}{dQ} \left( \frac{f(Q)}{\sqrt{-Q}} \right)\, ,
\end{align}
the solution is given by 
\begin{align}
\label{arbQ}
f(Q)= \sqrt{-Q} \int^Q dq \frac{\rho(q)}{\sqrt{ - q}} \, .
\end{align}
By using (\ref{arbQ}), one can find $f(Q)$ gravity corresponding to the arbitrary expansion history of the universe given by $a=a(t)$. 

At the end of the inflation, the matter could be generated by the quantum corrections, which are not included in the classical action. 
We now effectively include the effects by modifying (\ref{vQ_FLRW0}) as follows, 
\begin{align}
\label{vQ_FLRW0_eff}
\frac{f(Q)}{2} + 6 H^2 f'(Q) =&\, \rho_\mathrm{eff} \equiv \rho + \mathcal{J}_\rho (Q) \, , \nonumber \\
 - \left\{ \frac{f(Q)}{2} + 2 a^{-3} \frac{d}{dt} \left( a^3 H f'(Q) \right)\right\} =&\, p_\mathrm{eff} \equiv p + \mathcal{J}_p(Q)\, \, .
\end{align}
Here $\mathcal{J}_\rho(Q)$ and $\mathcal{J}_p(Q)$ are some functions of $Q$. 
Eq.~(\ref{cons2}) tells that $\rho_\mathrm{eff}$ and $p_\mathrm{eff}$ satisfy the conservation law as in (\ref{cons}), 
which gives, 
\begin{align}
\label{conseff}
\dot\rho + 3 H \left( \rho + p \right) = J \equiv - \dot Q \mathcal{J}_\rho'(Q) - 3 H \left( \mathcal{J}_\rho(Q) + \mathcal{J}_p(Q) \right) \, .
\end{align}
Then $J$ acts as a source of matter. 
We may choose $J$ so that $J$ does not vanish only in the period of the end of the inflation and generates matter. 
In general, the functions $\mathcal{J}_\rho$ and $\mathcal{J}_p$ can depend on $\dot Q$, $\ddot Q$, $\cdots$ in addition to $Q$ but 
just for simplicity, we choose $\mathcal{J}_\rho$ and $\mathcal{J}_p$ to be functions of only $Q$.

\subsection{Reconstruction of mimetic $f(Q)$ gravity}

Let us consider the mimetic $f(Q)$ gravity, where the scalar field $\phi$ couple with gravity in the following way, 
\begin{align}
\label{ActionQB}
S=\int d^4 x \sqrt{-g} \left\{ f(Q) + \lambda \left( g^{\mu\nu} \partial_\mu \phi \partial_\nu \phi + 1 \right) - V(\phi) + \mathcal{L}_\mathrm{matter} 
\right\}\, .
\end{align}
Here $\mathcal{L}_\mathrm{matter}$ is the Lagrangian density of matter and $\lambda$ is a Lagrange multiplier field.
The variation of the action with respect to $\lambda$ gives the mimetic constraint, 
\begin{align}
\label{mimeticconstraint}
0= g^{\mu\nu} \partial_\mu \phi \partial_\nu \phi + 1 \, . 
\end{align}
In the FLRW spacetime (\ref{FLRW}), a solution is given by 
\begin{align}
\label{mimetics}
\phi=t \, .
\end{align}
On the other hand, the first equation corresponding to the equations in (\ref{vQ_FLRW0}) is given by, 
\begin{align}
\label{vQ_FLRW0C0}
 - \rho - 2\lambda - V =&\, - \frac{f(Q)}{2} - 6 H^2 f'(Q) \, , \nonumber \\
 - p - 2\lambda + V =&\, \frac{f(Q)}{2} + 6H^2 f'(Q) + 2 \dot H f'(Q) - 24 H^2 \dot H f''(Q) \, ,
\end{align}
which can be solved with respect to $\lambda$ and $V$ as follows, 
\begin{align}
\label{fQmm1}
\lambda =&\, \frac{1}{4} \left( - \rho - p - 2 \dot H f'(Q) + 24 H^2 \dot H f''(Q) \right) \, , \nonumber \\
V =&\, \frac{1}{2} \left( -\rho + p + f(Q) + 12 H^2 f'(Q) + 2 \dot H f'(Q) - 24 H^2 \dot H f''(Q) \right) \, .
\end{align}
Here we have used (\ref{mimetics}). 
When $a=a(t)$ is given, 
$\rho$, $H$, and therefore $Q$ are expressed as a function of $t$. 
Then the first equation in (\ref{fQmm1}) gives the time-dependence of $\lambda$ as a solution. 
On the other hand, the second equation gives the $\phi$ dependence of $V(\phi)$ for arbitrary $f(Q)$, 
\begin{align}
\label{vQ_FLRW0C}
V \left( \phi \right) = \frac{1}{2} \left[ -\rho (t) + p (t) + f\left(Q\left(t\right) \right) + 12 H(t)^2 f' \left(Q\left(t\right)\right) 
+ 2 \dot H(t) f' \left(Q\left(t\right) \right) - 24 H(t)^2 \dot H(t) f'' \left(Q\left(t \right) \right) \right]_{t=\phi} \, .
\end{align}
Inversely, if we choose $V \left( \phi \right)$ by (\ref{vQ_FLRW0C}), we obtain 
the time-development of the universe expansion expressed by $a=a(t)$ as a solution. 

\section{Cosmological applications\label{SecV}}

In this section, we apply the formulations of the reconstruction in the last section to cosmology and we propose models which describe inflation, dark energy, 
and the unification of inflation and dark energy.

\subsection{A model of inflation}

We now consider the inflation by using Eq.(\ref{arbQ}). 
As after the inflation, radiation dominates, we consider only radiation as matter. 
In the radiation-dominated universe, the Hubble rate $H$ behaves as $H\sim \frac{1}{2t}$ and during the inflation, of course, $H=H_0$ $\left(H_0:\ \mbox{constant}\right)$. 
A unifying behaviour is given by 
\begin{align}
\label{Hex1b}
H(t) = \frac{H_0}{1 + \alpha \ln \left( 1 + \e^{\frac{2 H_0}{\alpha} \left( t - t_0 \right)} \right)} \, .
\end{align}
Here $\alpha$ is a positive constant and $t_0$ is a constant which corresponds to the time when the inflation ends. 
When $t\ll t_0$, we get $H$ becomes a constant $H\to H_0$, which corresponds to the inflation. 
On the other hand, when $t\gg t_0$, we find $H\to \frac{1}{2\left(t - t_0 \right)}$, which corresponds to the radiation-dominated universe. 
Because $Q=-{ 6}H^2$, one obtains 
\begin{align}
\label{Hex1b2}
\e^{\frac{2 H_0}{\alpha} \left( t - t_0 \right)} = \e^{\frac{1}{\alpha} \left(\sqrt{- \frac{{ 6}{H_0}^2}{Q}} - 1\right)} - 1 \quad \mbox{or} \quad 
t=t_0 + \frac{\alpha}{2H_0} \ln \left( \e^{\frac{1}{\alpha} \left( \sqrt{- \frac{{ 6}{H_0}^2}{Q}} - 1\right) } - 1 \right) \, .
\end{align}
Therefore $t$ is explicitly given as a function of $Q$. 

We now find 
\begin{align}
\label{derivatives}
\dot H = - \frac{2 H^2}{1 + \e^{-\frac{2 H_0}{\alpha} \left( t - t_0 \right)}} 
= 2 H^2 \left( 1 - \e^{-\frac{1}{\alpha} \left(\frac{H_0}{H} - 1 \right)} \right)
= { - \frac{Q}{3}} \left( 1 - \e^{-\frac{1}{\alpha} \left(\sqrt{- \frac{{ 6}{H_0}^2}{Q}}-1\right)} \right)\, , 
\end{align}
which will be used later. 

As it was mentioned around Eq.~(\ref{conseff}), we may assume the radiation is generated at the end of the inflation $t\sim t_0$. 
One may assume the behaviour of the energy density $\rho$ of the radiation as follows, { 
\begin{align}
\label{Hex1b3}
\rho = - \frac{Q \left( Q + 6 {H_0}^2 \right)}{2\kappa^2 \left( - Q + 6 {H_0}^2 \right)} 
= - \frac{1}{2\kappa^2} \left( - Q - 12 {H_0}^2 + \frac{72{H_0}^4}{ - Q + 6{H_0}^2} \right) \, .
\end{align}}
When $t\ll t_0$, because $Q= -{ 6}H^2 \to - { 6}{H_0}^2$, we obtain $\rho \to 0$. 
On the other hand, when $t\gg t_0$, because $Q\ll { 6}{H_0}^2$, $\rho$ behaves as $\rho\to - \frac{3Q}{{ 6}\kappa^2} = \frac{3H^2}{\kappa^2}$ as 
expected in Einstein's gravity. 
The equation of state (EoS) parameter of the radiation is $\frac{1}{3}$, the pressure $p$ is given by { 
\begin{align}
\label{Hex1b3p}
p = - \frac{Q \left( Q + 6{H_0}^2 \right)}{6 \kappa^2 \left( - Q + 6{H_0}^2 \right)} \, .
\end{align}}
Note that the above $\rho$ and $p$ do not satisfy the conservation law (\ref{cons}) but violate it as in (\ref{conseff}). 
In fact, by using (\ref{derivatives}), we find, { 
\begin{align}
\label{consmod1}
J=&\, \dot \rho + 3 H \left( \rho + p \right) 
= \frac{H}{\kappa^2} \left\{ \frac{ 2 Q \left( - Q^2 + 12 {H_0}^2 Q + 36 {H_0}^4 \right)}{ \left( - Q + 6{H_0}^2\right)^2 } 
\left( 1 - \e^{-\frac{1}{\alpha} \left( \sqrt{- \frac{6{H_0}^2}{Q}}-1\right)} \right) 
 - \frac{Q \left( Q + 6{H_0}^2 \right)}{ - Q + 6{H_0}^2 } \right\} \, ,
\end{align}}
which vanishes at the early time $t\ll t_0$ or $Q\to - { 6}{H_0}^2$ and at the late time $t\gg t_0$ or $Q\to 0$. 
The source $J$ vanishes except only at the end of the inflation $t\sim t_0$, when the radiation is generated. 
There are some ambiguities in choosing $\mathcal{J}_\rho(Q)$ in (\ref{vQ_FLRW0_eff}). 
For simplicity, one may choose $\mathcal{J}_\rho(Q)=0$. 
Then Eq.~(\ref{conseff}) gives, { 
\begin{align}
\label{fp}
\mathcal{J}_p = - \frac{1}{3\kappa^2} \left\{ \frac{ 2 Q \left( - Q^2 + 12 {H_0}^2 Q + 36 {H_0}^4 \right)}{ \left( - Q + 6{H_0}^2\right)^2 } 
\left( 1 - \e^{-\frac{1}{\alpha} \left( \sqrt{- \frac{6{H_0}^2}{Q}}-1\right)} \right) 
 - \frac{Q \left( Q + 6{H_0}^2 \right)}{ - Q + 6{H_0}^2 } \right\} \, ,
\end{align}}
which may effectively express the quantum generation of the radiation. 

By using (\ref{arbQ}), we find the explicit form of $f(Q)$ { 
\begin{align}
\label{arbQinf}
f(Q)= \frac{1}{\kappa^2} \left\{ \frac{Q^2}{3} + 12 {H_0}^2 Q + {H_0}^3 \sqrt{-\frac{2}{3}Q} 
\mathrm{Arctan} \left(\sqrt{-\frac{Q}{6{H_0}^2}}\right) + C \sqrt{-Q} \right\} \, .
\end{align}}
Here $C$ is a constant of the integration
Note that we have chosen $\rho_\mathrm{eff}=\rho$. 

Let us define the slow-roll parameters as follows, 
\begin{align}
\label{slowrole}
\epsilon_1 = - \frac{\dot H}{H^2}\, , \quad 
\epsilon_{n+1} = \frac{\dot \epsilon_n}{H \epsilon_n}\, , \quad \left(n=1,2,3,\cdots \right)\, .
\end{align}
Then the scalar spectral index $n_s$ and the tensor-to-scalar ratio $r$ are given by 
\begin{align}
\label{index0}
n_s = \left. \left[1 -2 \epsilon_1 - 2 \epsilon_2 \right] \right|_\mathrm{h.c.} \, , \quad r = \left. 16 \epsilon_1 \right|_\mathrm{h.c.} \, .
\end{align}
where the suffix `$\mathrm{h.c.}$' symbolises the horizon crossing instant of the CMB scale mode in which we are interested. 
Planck 2018 results constrain the observational indices as follows, 
\begin{align}
\label{index}
n_s = 0.9649 \pm 0.0042\, , \quad r < 0.064\, .
\end{align}
We now consider how these constraints can be satisfied. 

For the model (\ref{Hex1b}), by using (\ref{derivatives}), we find 
\begin{align}
\label{slowrole2}
\epsilon_1 = \frac{2}{1 + \e^{- \frac{2 H_0}{\alpha} \left( t - t_0 \right)}} \, , \quad 
\epsilon_2 = \frac{2\left\{ 1 + \alpha \ln \left( 1 + \e^{\frac{2 H_0}{\alpha} \left( t - t_0 \right)} \right) \right\}
\e^{- \frac{2 H_0}{\alpha} \left( t - t_0 \right)}}{\alpha \left( 1 + \e^{- \frac{2 H_0}{\alpha} \left( t - t_0 \right)}\right)}\, ,
\end{align}
which give 
\begin{align}
\label{index00}
n_s = 1 - \frac{4\left[ 1 + \frac{1}{\alpha} \left\{ 1 + \alpha \ln \left( 1 + \e^{\frac{2 H_0}{\alpha} \left( t_\mathrm{h.c.} - t_0 \right)} \right) \right\}
\e^{-\frac{2 H_0}{\alpha} \left( t_\mathrm{h.c.} - t_0 \right)}\right]}
{1 + \e^{- \frac{2 H_0}{\alpha} \left( t_\mathrm{h.c.} - t_0 \right)}} \, , \quad 
r = \frac{32}{1 + \e^{-\frac{2 H_0}{\alpha} \left( t_\mathrm{h.c.} - t_0 \right)}} \, .
\end{align}
Here $t_\mathrm{h.c.}$ is the time of the horizon crossing. 
The second constraint about $r$ in (\ref{index}) gives
\begin{align}
\label{estimate1}
\e^{-\frac{2 H_0}{\alpha} \left( t_\mathrm{h.c.} - t_0 \right)}>500\, .
\end{align}
Therefore the second term of $n_s$ in (\ref{index}) should be 
\begin{align}
\label{estimate2}
2\epsilon_1 = \frac{4}{1 + \e^{-\frac{2 H_0}{\alpha} \left( t_\mathrm{h.c.} - t_0 \right)}} < 0.008 \, .
\end{align}
On the other hand, the third term of $n_s$ (\ref{index}) could be given by 
\begin{align}
\label{estimate3}
2\epsilon_2 = \frac{4\left\{ 1 + \alpha \ln \left( 1 + \e^{\frac{2 H_0}{\alpha} \left( t - t_0 \right)} \right) \right\}
\e^{- \frac{2 H_0}{\alpha} \left( t - t_0 \right)}}{\alpha \left( 1 + \e^{- \frac{2 H_0}{\alpha} \left( t - t_0 \right)}\right)} 
\sim \frac{4}{\alpha} \, .
\end{align}
If we assume, 
\begin{align}
\label{estimate4}
2\epsilon_2 \sim 1 - n_s \sim 0.035 \, ,
\end{align}
one gets 
\begin{align}
\label{estimate5}
\alpha \sim 1.1\times 10^2 \, .
\end{align}
Therefore the constraints in (\ref{index}) are satisfied. 

\subsection{Dark energy epoch}

Let us now consider a model that describes the dark energy epoch using Eq.(\ref{arbQ}). 
Here just for simplicity, we consider the model mimicking the $\Lambda$CDM model. 

The scale factor $a(t)$ in the $\Lambda$CDM model is given by 
\begin{align}
\label{LCDM1}
a(t) = a_0 \sinh^\frac{2}{3} \left( \alpha t \right)\, ,
\end{align}
where $a_0$ and $\alpha$ are positive constants. 
Eq.~(\ref{LCDM1}) gives the following Hubble rate $H(t)$, 
\begin{align}
\label{LCDM2}
H(t) = \frac{2}{3}\alpha \coth \left( \alpha t \right)\, .
\end{align}
Therefore we find 
\begin{align}
\label{LCDM3}
Q(t) = - { \frac{8}{3}}\alpha^2 \coth^2 \left( \alpha t \right) 
= - { \frac{8}{3}}\alpha^2 \frac{\sinh^2 \left( \alpha t \right) + 1}{\sinh^2 \left( \alpha t \right) }\, , 
\end{align}
which gives 
\begin{align}
\label{LCDM4}
a(t) = a_0 \left( Q(t) + { \frac{8}{3}}\alpha^2 \right)^{-\frac{1}{3}} \, .
\end{align}
As a matter, one may consider the baryonic matter, whose EoS is $\frac{1}{3}$. 
Then the energy density $\rho$ behaves as 
\begin{align}
\label{LCDM5}
\rho = \rho_0 a(t)^{-3} = \frac{\rho_0}{{a_0}^3}\left( Q(t) + { \frac{8}{3}}\alpha^2 \right) \, .
\end{align}
The energy density $\rho$ is conserved in the dark energy epoch, we put $\mathcal{J}_\rho (Q)=\mathcal{J}_p(Q)=0$ in (\ref{vQ_FLRW0_eff}). 
Then by using (\ref{arbQ}), we get the explicit form of $f(Q)$ as follows, 
\begin{align}
\label{arbQB}
f(Q)= \frac{2\rho_0}{3{a_0}^3} \left( Q^2 + \frac{ 2}{3} \alpha^2 Q + \frac{3}{2} C \sqrt{-Q} \right) \, .
\end{align}
Here $C$ is a constant of the integration, again. 

In summary, we can realise the $F(Q)$ gravity cosmology just like in the $\Lambda$CDM model without the cosmological constant and the real dark matter. 
The dark matter is, of course, necessary for the structure formation etc. 
In this subsection, just for simplicity, we considered a toy model without real dark matter. 

\subsection{Unification of Inflation and Dark Energy Epochs}

Let us now consider a model which describes both the inflation and dark energy epochs by using the Eq.(\ref{arbQ}), again. 
Note that such unification has been achieved in $f(R)$ gravity for several realistic theories, see Refs.~{Nojiri:2003ft, Nojiri:2010wj}. 
In this section, we use the $e$-folding number $N$ defined by $a=\e^N$ instead of the cosmological time, which makes the treatments of the matter comprehensible. 

\subsubsection{Models of unification and early dark energy}

As in (\ref{Hex1b3}), we assume that the matter was generated at the end of the inflation and the energy density $\rho$ is given by 
\begin{align}
\label{Uni1}
\rho(N) = \frac{\e^{n \left( N - N_0\right) }}{1 + \e^{n\left( N - N_0\right) }} \left( \rho_0^\mathrm{radiation} \e^{-4N} + \rho_0^\mathrm{baryon} \e^{-3N} \right) \, .
\end{align}
Here $\rho_0^\mathrm{radiation}$ and $\rho_0^\mathrm{baryon}$ are positive constants and $n$ is also a constant larger than $4$. 
In addition to the radiation corresponding to the first term $\rho_0^\mathrm{radiation} \e^{-4N}$ in the parentheses of (\ref{Uni1}), 
we include the baryon matter corresponding to the second term $rho_0^\mathrm{baryon} \e^{-3N}$. 
The factor $\frac{\e^{n\left( N-N_0\right) }}{1 + \e^{n\left( N-N_0\right) }}$ expresses the creation of matter 
and $N_0$ corresponds to the $e$-folding number when the inflation ends. 
In fact, when $N\ll N_0$, one finds $\frac{\e^{n\left( N-N_0\right) }}{1 + \e^{n\left(N-N_0\right) }} \sim \e^{n \left( N - N_0 \right)} \to 0$ and therefore $\rho(N)\to 0$, 
and when $N\gg N_0$, $\frac{\e^{N-N_0}}{1 + \e^{N-N_0}}\to 1$ and $\rho(N) \to \rho_0^\mathrm{radiation} \e^{-4N} + \rho_0^\mathrm{baryon} \e^{-3N}$. 

We now consider the model, where $Q$ and therefore $H$ is given by, 
\begin{align}
\label{Uni2}
Q =&\, -{ 6}H^2 = - \frac{ { 6}{H_0}^2 \left( 1 + \epsilon \e^{2 N - 2N_0} \right)}{1 + \e^{2 N - 2 N_0}} - { 2\kappa^2}\rho(N) \, .
\end{align}
Here $H_0$ and $\epsilon$ are positive constants. 
The first term in the r.h.s. behaves as $- \frac{ { 6}{H_0}^2 \left( 1 + \epsilon \e^{N-N_0} \right)}{1 + \e^{N-N_0}} \to - { 6}{H_0}^2$ when $N\ll N_0$, 
which may play the role of the large effective cosmological constant generating inflation. 
It also behaves as $- \frac{ { 6}{H_0}^2 \left( 1 + \epsilon \e^{N-N_0} \right)}{1 + \e^{N-N_0}} \to -{ 6}\epsilon{H_0}^2$ when $N\gg N_0$, 
which may correspond to the small effective cosmological constant generating the late-time accelerating expansion 
if we choose $\epsilon$ very small. 

One may obtain $N$ as a function of $Q$, $N=N(Q)$ from Eq.~(\ref{Uni2}) although it is difficult to solve Eq.~(\ref{Uni2}) explicitly. 
$N$ as a function of $Q$, $N=N(Q)$ although the explicit form is very complicated and we do not write down the corresponding explicit form. 
By using the expression $N=N(Q)$, we may write $\rho$ in (\ref{Uni1}) as a function of $Q$, $\rho=\rho(Q)$. 
By putting $\mathcal{J}_\rho (Q)=0$ in (\ref{vQ_FLRW0_eff}) and by using (\ref{arbQ}), one may find the form of $f(Q)$, again. 

Note that it is not so difficult to include the early dark energy. 
As a simple model, we modify (\ref{Uni2}) as follows, 
\begin{align}
\label{Uni3}
Q =&\, -{ 6}H^2 = - \frac{ { 6}{H_0}^2 \left\{ 1 - \epsilon \e^{-2N_0 + 2 N_\mathrm{EDE}} + \epsilon \e^{-2N_0} \left( \e^N - \e^{N_\mathrm{EDE}} \right)^2 \right\}}
{1 - \e^{-2N_0 + 2 N_\mathrm{EDE}}+ \e^{ - 2 N_0} \left( \e^N - \e^{N_\mathrm{EDE}} \right)^2} - { 2\kappa^2}\rho(N) \, .
\end{align}
Here $N_\mathrm{EDE}$ corresponds to the early dark energy epoch when the redshift is given by $z\sim 3000$. 
Therefore one gets $N_{EDE} \ll N_0$. 
When $N\ll N_0$, the first term in (\ref{Uni3}) goes to a constant $- \frac{ { 6}{H_0}^2 \left\{ 1 - \epsilon \e^{-2N_0 + 2 N_\mathrm{EDE}} 
+ \epsilon \e^{-2N_0} \left( \e^N - \e^{N_\mathrm{EDE}} \right)^2 \right\}}
{1 - \e^{-2N_0 + 2 N_\mathrm{EDE}}+ \e^{ - 2 N_0} \left( \e^N - \e^{N_\mathrm{EDE}} \right)^2} \to { 6}{H_0}^2$, 
which corresponds to the effective cosmological constant generating inflation. 
When $N\gg N_\mathrm{EDE}$, we find $- \frac{ { 6}{H_0}^2 \left\{ 1 - \epsilon \e^{-2N_0 + 2 N_\mathrm{EDE}} + \epsilon \e^{-2N_0} \left( \e^N - \e^{N_\mathrm{EDE}} \right)^2 \right\}}
{1 - \e^{-2N_0 + 2 N_\mathrm{EDE}}+ \e^{ - 2 N_0} \left( \e^N - \e^{N_\mathrm{EDE}} \right)^2} \to { 6} \epsilon {H_0}^2$, which corresponds to the dark energy at the late universe. 
We should note that the first term in (\ref{Uni3}) has a stationary point at $N\sim N_\mathrm{EDE}$ as 
$- \frac{ { 6}{H_0}^2 \left\{ 1 - \epsilon \e^{-2N_0 + 2 N_\mathrm{EDE}} + \epsilon \e^{-2N_0} \left( \e^N - \e^{N_\mathrm{EDE}} \right)^2 \right\}}
{1 - \e^{-2N_0 + 2 N_\mathrm{EDE}}+ \e^{ - 2 N_0} \left( \e^N - \e^{N_\mathrm{EDE}} \right)^2} \to
 - \frac{ { 6}{H_0}^2 \left\{ 1 - \epsilon \e^{-2N_0 + 2 N_\mathrm{EDE}} \right\}}{1 - \e^{-2N_0 + 2 N_\mathrm{EDE}}}$, 
which may correspond to the early dark energy. 
In summary, we presented $f(Q)$ gravity which unifies inflation with dark energy even with account of early dark energy.

\subsubsection{Asymptotic behaviours}

Let us now estimate the asymptotic behaviour of $f(Q)$ in Eq.~(\ref{Uni3}). 
We also choose $n\gg 4$ in (\ref{Uni1}). 
\begin{itemize}
\item When $N\ll N_0$, Eq.~(\ref{Uni3}) behaves as 
\begin{align}
\label{Uni4}
Q \sim - { 6}{H_0}^2 \left( 1 + 2 \e^{-2N_0 + N_\mathrm{EDE} + N} \right) \, .
\end{align}
Here it is also assumed $\epsilon\ll 1$. 
Eq.~(\ref{Uni4}) can be solved with respect to $N$ as follows, 
\begin{align}
\label{Uni5}
\e^{N}= - \frac{\e^{2N_0 - N_\mathrm{EDE}}}{2} \left( \frac{Q}{{ 6}{H_0}^2} + 1 \right) \, ,
\end{align}
and (\ref{Uni1}) behaves as 
\begin{align}
\label{Uni6}
\rho(N) \sim \rho_0^\mathrm{radiation} \e^{- N_0} \e^{\left( n - 4 \right) N} 
\sim \rho_0^\mathrm{radiation} \e^{- N_0} \left\{ - \frac{\e^{2N_0 - N_\mathrm{EDE}}}{2} \right\}^{n-4} \left( \frac{Q}{{ 6}{H_0}^2} + 1 \right)^{n-4} \, .
\end{align}
Then Eq.~(\ref{arbQ}) gives, 
\begin{align}
\label{Uni7}
f(Q) \sim \rho_0^\mathrm{radiation} \e^{- N_0} \left\{ - \frac{\e^{2N_0 - N_\mathrm{EDE}}}{2} \right\}^{n-4} \sqrt{-Q} \int^Q \frac{dq}{\sqrt{ - q}} 
 \left( \frac{q}{{ 6}{H_0}^2} + 1 \right)^{n-4} 
\, .
\end{align}
Especially when $n=5$, we obtain 
\begin{align}
\label{Uni8}
f(Q) \sim \rho_0^\mathrm{radiation} \e^{- N_0} \left\{ - \frac{\e^{2N_0 - N_\mathrm{EDE}}}{2} \right\} 
\left( \frac{Q^2}{{ 9}{H_0}^2} + 2 Q + C_1 \sqrt{-Q}\right) \, .
\end{align}
Here $C_1$ is a constant of the integration. 
\item When $N\gg N_\mathrm{EDE}$, Eq.~(\ref{Uni3}) gives, 
\begin{align}
\label{Uni9}
Q \sim -{ 6}\epsilon{H_0}^2 \left( 1 + \frac{\e^{2N_0}}{\epsilon} \e^{-2N} \right) \, .
\end{align}
that is, 
\begin{align}
\label{Uni10}
\e^{-2N} \sim - \epsilon \e^{-2N_0} \left( \frac{Q}{{ 6}\epsilon {H_0}^2} + 1 \right)\, ,
\end{align}
which gives, 
\begin{align}
\label{Uni11}
\rho(Q) \sim \rho_0^\mathrm{baryon} \e^{-3N} = \rho_0^\mathrm{baryon} \epsilon^\frac{3}{2} \e^{-3N_0}\left( - \frac{Q}{{ 6}\epsilon {H_0}^2} - 1 \right)^\frac{3}{2} \, .
\end{align}
and 
Eq.~(\ref{arbQ}) gives, { 
\begin{align}
\label{Uni12}
f(Q) \sim&\, \rho_0^\mathrm{baryon} \epsilon^\frac{3}{2} \e^{-3N_0} \epsilon {H_0}^2 \sqrt{-\frac{Q}{6\epsilon {H_0}^2}} \nonumber \\
& \times 
\left[ \sqrt{ - 1 - \frac{Q}{6\epsilon {H_0}^2}} \left\{ \frac{1}{2} \left( - \frac{Q}{6\epsilon {H_0}^2} \right)^\frac{3}{2} - \frac{5}{4}\sqrt{- \frac{Q}{6\epsilon {H_0}^2}} \right\}
+ \frac{3}{4} \cosh^{-1} \left( \sqrt{- \frac{Q}{6\epsilon {H_0}^2}} \right) + C \right]\, .
\end{align}}
Here $\cosh^{-1}$ is the inverse function of $\cosh$ $\left( y=\cosh x \, , \ x=\cosh^{-1} y \right)$ and $C$ is a constant of the integration. 
\end{itemize}
Therefore, we presented asymptotics of $f(Q)$ which unifies the inflation with dark energy

\subsection{Models of mimetic $f(Q)$ gravity}

In this section, for completeness, we construct mimetic $f(Q)$ gravity by using the formula in (\ref{vQ_FLRW0C}) so that it unifies the inflation with dark energy.
For simplicity, in this subsection, we choose $f(Q)$ as follows, 
\begin{align}
\label{sqrtQ1}
f(Q) = \frac{Q}{2\kappa^2} + f_1 Q^2 \, .
\end{align}
Here $f_1$ is a constant. 

\subsubsection{A mimetic model of inflation}

In the case of the inflation epoch, 
we assume (\ref{Hex1b}), (\ref{Hex1b3}), and also (\ref{Hex1b3p}). 
Then Eq.~(\ref{vQ_FLRW0C}) gives { 
\begin{align}
\label{vQ_FLRW0Cinf}
V \left( \phi \right) =&\, - \frac{{H_0}^2 \left\{ \left( 1 + \alpha \ln \left( 1 + \e^{\frac{2 H_0}{\alpha} \left( \phi - t_0 \right)} \right) \right)^2 - 1\right\}
}{\kappa^2 \left( 1 + \alpha \ln \left( 1 + \e^{\frac{2 H_0}{\alpha} \left( \phi - t_0 \right)} \right) \right)^2
\left\{ \left( 1 + \alpha \ln \left( 1 + \e^{\frac{2 H_0}{\alpha} \left( \phi - t_0 \right)} \right) \right)^2 + 1\right\}} \nonumber \\
&\, + \left( 3 - \frac{2}{1 + \e^{-\frac{2 H_0}{\alpha} \left( \phi - t_0 \right)}} \right) 
\left\{ \frac{{H_0}^2}{\kappa^2\left(1 + \alpha \ln \left( 1 + \e^{\frac{2 H_0}{\alpha} \left( \phi - t_0 \right)} \right)\right)^2}
 - \frac{36 f_1{H_0}^4}{\left(1 + \alpha \ln \left( 1 + \e^{\frac{2 H_0}{\alpha} \left( \phi - t_0 \right)} \right)\right)^4} \right\} \, .
\end{align}}
Hence, we find that even in the mimetic theory, we can reconstruct a model which describes inflation straightforwardly. 

\subsubsection{A mimetic model of dark energy}

Similarly, at the dark energy epoch, the model (\ref{LCDM2}) mimicking the $\Lambda$CDM model with (\ref{LCDM5}) is { 
\begin{align}
\label{DEmimetic1}
V(\phi)= \frac{4 \rho_0 \alpha^2}{3{a_0}^3 \sinh^2 \left( \alpha \phi \right)} 
+ \frac{4\alpha^2}{3\kappa^2} 
+ \frac{64}{9} \alpha^4 f_1 \left( - \coth^4 \left( \alpha \phi \right) 
+ \frac{3\coth^2 \left( \alpha \phi \right)}{\sinh^2 \left( \alpha \phi \right)} \right) \, .
\end{align}}
Then we obtain an explicit model, again.

\subsubsection{A mimetic model unifying inflation and dark energy}

When we consider the unification of the inflation epoch and the dark energy epoch in (\ref{Uni2}) and (\ref{Uni3}), we used the $e$-folding $N$, 
which makes it difficult to use the formula (\ref{vQ_FLRW0C}) written in terms of the cosmological time $t$ and therefore we cannot identify the scalar field $\phi$ with 
the $e$-folding number $N$. 
One way to avoid this problem is to choose a coordinate system where we can identify $N$ as a time coordinate, instead of the standard FLRW one in (\ref{FLRW}), 
as follows,
\begin{align}
\label{Ncoord}
ds^2 = - \alpha(N)^2 dN^2 + \e^{2N}\sum_{i=1,2,3} \left( dx^i \right)^2\, .
\end{align}
Here $\alpha$ is a function of $N$ which describes the expansion of the universe. 
Then the mimetic constraint (\ref{mimeticconstraint}) gives $\phi=N$ instead of (\ref{mimetics}). 
Eq.~(\ref{Ncoord}) also shows
\begin{align}
\label{NH}
\alpha= \frac{dt}{dN}=\frac{1}{H}\, .
\end{align}
Anyway, because $\frac{DN}{dt}=H$, we can express $t$ as a function of $N$, 
\begin{align}
\label{tN}
t(N)=\int \frac{dt}{dN} dN = \int \frac{dN}{H(N)}\, ,
\end{align}
which could be solved with respect to $N$ as $N=N(t)$. 


We now investigate the asymptotic behaviours of $V(\phi)$ in Eq.~(\ref{vQ_FLRW0C}) for the model of Eq.~(\ref{Uni3}) 
with $n\gg 4$ in (\ref{Uni1}). 
\begin{itemize}
\item When $N\ll N_0$, the Hubble rate $H$ behaves as 
\begin{align}
\label{AsHforNlrg}
H \sim H_0 \left( 1 + \e^{-2N_0 + N_\mathrm{EDE} + N} \right) \, .
\end{align}
Then by using (\ref{tN}), we find 
\begin{align}
\label{AsHforNnglrg}
t \sim \frac{1}{H_0} \int dN \left( 1 - \e^{-2N_0 + N_\mathrm{EDE} + N} \right) 
\sim \frac{N}{H_0} \, .
\end{align}
therefore as expected from the asymptotically de Sitter behavior, we find $N\sim H_0 t$. 
Eq.~(\ref{Uni4}) also tells
\begin{align}
\label{Uni4min}
Q \sim - { 6}{H_0}^2 \left( 1 + 2 \e^{-2N_0 + N_\mathrm{EDE} + H_0 t} \right) \, .
\end{align}
Here we neglect the integration constant because we assume $t$ is negative and large. 
Then Eq.~(\ref{vQ_FLRW0C}) shows { 
\begin{align}
\label{vQ_FLRW0CNnglrg}
V \left( \phi \right) \sim 
\frac{{H_0}^2}{\kappa^2} \left( \frac{3}{2} + 4 \e^{-2N_0 + N_\mathrm{EDE} + H_0 \phi} \right)
 - f_1 {H_0}^4 \left( 54 + 360 \e^{-2N_0 + N_\mathrm{EDE} + H_0 \phi} \right) \, .
\end{align}}
\item On the other hand, when $N\gg N_\mathrm{EDE}$, 
the Hubble rate $H$ behaves as 
\begin{align}
\label{AsHforNpslrg}
H \sim \sqrt\epsilon H_0 \left( 1 + \frac{\e^{2N_0}}{2\epsilon} \e^{-2N} \right) \, .
\end{align}
Then one gets $N\sim \sqrt\epsilon H_0 t$ as we find in (\ref{AsHforNnglrg}). 
We neglect the integration constant because $t$ could be positive and large. 
Then Eq.~(\ref{Uni9}) gives, 
\begin{align}
\label{Uni9mim}
Q \sim -{ 6}\epsilon{H_0}^2 \left( 1 + \frac{\e^{2N_0}}{\epsilon} \e^{-2 \sqrt\epsilon H_0 t} \right) \, .
\end{align}
Finally, we find { 
\begin{align}
\label{vQ_FLRW0Cpslrg}
V \left( \phi \right) \sim 
\frac{\epsilon {H_0}^2}{\kappa^2} \left( \frac{3}{2} + \frac{\e^{2N_0}}{2\epsilon} \e^{-2 \sqrt\epsilon H_0 \phi } \right) 
 - 108 \epsilon^2 {H_0}^4 f_1 \, .
\end{align}}
\end{itemize}
Hence, the unification of inflation with dark energy in mimetic $f(Q)$ is possible.

\section{Summary and discussion\label{SecVI}}

In this paper, after a brief review of $f(Q)$ gravity, we proposed a consistent formulation of the $f(Q)$ gravity by using the metric 
and only the four scalar fields which give the connection as in (\ref{G1B}). 
Then the equations given by the variation of the action with respect to these fields become well-defined. 
We gave the corresponding quantities in (\ref{eq1}) and (\ref{eq3}) which were used when we considered the consistent equations. 
We calculated the explicit forms of the above quantities in the FLRW spacetime as in (\ref{Xa}) and (\ref{Einstein}) and we found 
that the coincident gauge is not consistent in the FLRW spacetime as shown in (\ref{gammaH}). 
By using the quantities in the FLRW spacetime, we proposed simple formulae of the reconstruction in the $f(Q)$ gravity in (\ref{arbQ}) 
and its mimetic extension in (\ref{vQ_FLRW0C}). 
Such a formulation makes it possible to construct a model realizing the inflation epoch (\ref{arbQinf}) corresponding to (\ref{Hex1b}). 
This model could give the scalar spectral index $n_s$ and the tensor-to-scalar ratio $r$ consistent with Planck 2018 data. 
It is also constructed a model describing the dark energy epoch as in (\ref{arbQB}) corresponding to the scale factor $a(t)$ in the $\Lambda$CDM theory (\ref{LCDM1}). 
The unification of the inflation epoch and the dark energy epoch was discussed by using a model in (\ref{Uni3}). 
The corresponding asymptotic behaviour of $f(Q)$ as in (\ref{Uni8}) and (\ref{Uni12}) is found. 

Let us now comment on the formulation of the cosmological reconstruction. 
The FLRW spacetime in (\ref{FLRW}) can be characterised by only one function $a(t)$, which is the scale factor, up to the coordinate transformation. 
This is one of the important reasons why we can reconstruct a model by using only one function $f(Q)$. 
Therefore it could be difficult to reconstruct models realising general spherically symmetric spacetime and more general spacetime only by $f(Q)$ gravity. 
Consider the case of the scalar-tensor theory with the action
\begin{align}
\label{st1}
S_\phi = \int d^4x \sqrt{-g} \left\{ \frac{1}{2} \sum_{a,b} A_{(ab)} \left( \phi^{(c)} \right) g^{\mu\nu} \partial_\mu \phi^{(a)} \partial_\nu \phi^{(b)} - V \left( \phi^{(c)} \right) 
\right\} \, .
\end{align}
Here we include $N$ scalar fields, $\phi^{(a)}$ $\left( a = 1,2,\cdots,N \right)$. 
Note that there is an ambiguity in the redefinition of the scalar fields $\phi^{(a)} \to {\tilde\phi}^{(a)}\left( \phi^{(b)} \right)$, which changes $A_{(ab)}$ by 
$A_{(ab)} \left( \phi^{(c)} \right) \to {\tilde A}_{(ab)} \left( {\tilde \phi}^{(c)} \right)\equiv 
\sum_{d,f} \frac{\partial \phi^{(d)}}{\partial {\tilde \phi}^{(a)}} {\partial \phi^{(f)}}{\partial {\tilde \phi}^{(b)}} A_{(d f)} \left( \phi^{(h)}\left({\tilde \phi}^{(c)}\right) \right)$, 
which is an analogue of the coordinate transformation. 

In the case of $N=1$, by the change of the variable, we may choose $A_{(ab)}$ to be a unity, $A_{(ab)}=1$, or a minus unity $A_{(ab)}=-1$. 
After that, there remains one function, that is, the potential $V(\phi)$, which is one of the main reasons why the one-scalar model can reconstruct the model realizing 
the general FLRW universe \cite{Nojiri:2005pu, Capozziello:2005tf}. 

For the spherically symmetric spacetime, the metric is given by 
\begin{align}
\label{ssst}
ds^2 = g_{tt} dt^2 + 2 g_{tr} dt dr + g_{rr} dr^2 + r^2 \left( d\theta^2 + \sin^2\theta d\phi^2 \right) \, .
\end{align}
Here we choose the radial coordinate $r$ to be a radius given by the angles, that is, the radius is defined so that the area of the fixed $r$ becomes $4\pi r^2$. 
Then there appear three functions $g_{tt}$, $g_{tr}=g_{rt}$, and $g_{rr}$. 
Note we can often eliminate $g_{tr}$ by the coordinate transformation of $t$. 
This demonstrates that we need at least two scalar fields to realise the spherically symmetric spacetime (\ref{ssst}) in the framework of 
the scalar-tensor theory \cite{Nojiri:2020blr}. 
In the two scalar theory, there appear three components of $A_{(ab)} \left( \phi^{(c)} \right)$ and potential $V \left( \phi^{(c)} \right)$. 
Because there is an ambiguity in the field redefinition, which has two functional degrees of freedom, there remain only two functional degrees of freedom 
in the four functional degrees $A_{(ab)} \left( \phi^{(c)} \right)$ and $V \left( \phi^{(c)} \right)$. 
Instead of using two scalar fields, we may include another function like the Gauss-Bonnet coupling \cite{Nashed:2021cfs}. 

In general four-dimensional spacetime, the metric has ten components and four gauge degrees of freedom corresponding to the coordinate transformations. 
This shows us that in order to realise the general spacetime in the framework of the scalar-tensor theory in (\ref{st1}), we need at least four scalar fields as shown 
in \cite{Nashed:2024jqw}. 
When $N=4$ in (\ref{st1}), there appear ten functions $A_{(ab)} \left( \phi^{(c)} \right)$ and potential $V \left( \phi^{(c)} \right)$ and there are four redundancies by 
the redefinition of the scalar fields. 
Therefore the scalar-tensor theory (\ref{st1}) has one extra functional degree of freedom. 
Hence, one may choose $V=0$. 
As a result, we obtain a non-linear $\sigma$ model, whose target space is a four-dimensional manifold. 

Interestingly, the reconstructed models often include ghosts. 
The kinetic energy of the ghosts is unbounded below in the classical theory and negative norm states appear in the quantum theory. 
Such ghosts are known as the Fadeev-Popov ghosts in the gauge theories \cite{Kugo:1979gm, Kugo:1977zq}. 
It has been, however, shown that ghosts can be eliminated by imposing constraints like the mimetic constraint (\ref{mimeticconstraint}). 
The obtained models are applied to several kinds of spacetimes (for example, see \cite{Nojiri:2023dvf}). 
In the same way, we may extend the formulation of the reconstruction in $f(Q)$ gravity for different types of spacetime, including spherically symmetric spacetimes. 
This will be considered elsewhere.

{ 
\section*{ACKNOWLEDGEMENTS}

This work was partially supported by MICINN (Spain), project PID2019-104397GB-I00 and by the program Unidad
de Excelencia Maria de Maeztu CEX2020-001058-M, Spain (S.D.O). 
}

\end{document}